\documentclass[doublecol]{epl2}
\usepackage{epsfig}
\usepackage{color}
\usepackage{amsmath}
\usepackage{amssymb}
\usepackage{hyperref}

\title{Autowaves in a dc complex plasma confined behind a de Laval nozzle}

\author{M. A. Fink$^1$, S. K. Zhdanov$^1$, M. Schwabe$^{1,2}$, M. H. Thoma$^1$, H. H\"ofner$^1$,
H. M. Thomas$^1$, and G. E. Morfill$^1$}

\institute{$^1$Max-Planck-Institute for Extraterrestrial Physics,
Giessenbachstr., 85748 Garching, Germany

$^2$Department of Chemical and Biomolecular Engineering, University
of California, Berkeley, CA 94720, USA}


\abstract{ Experiments to explore stability conditions and topology
of a dense microparticle cloud supported against gravity by a gas
flow were carried out. By using a nozzle shaped glass insert within
the glass tube of a dc discharge plasma chamber a
weakly ionized gas flow through a de Laval nozzle was produced. The experiments were performed using neon gas at a pressure of 100 Pa and melamine-formaldehyde
particles with a diameter of $3.43~\mu\text{m}$. The capturing and
stable global confining of the particles behind the nozzle in the
plasma were demonstrated. The particles inside the cloud
behaved as a single convection cell inhomogeneously structured along
the nozzle axis in a tube-like manner. The pulsed acceleration
localized in the very head of the cloud mediated by collective
plasma-particle interactions and the resulting wave pattern were
studied in detail.
}

\begin{document}

\maketitle

\section{Introduction}

Solid 'seeds', the dust- or micro-particles incorporated into a
weakly ionized gas, can fundamentally alternate the low frequency
dynamics of the entire gas-plasma-particles system. During the last
decades such kind of \emph{complex plasmas} has been of great
particular interest in complex system studies
\cite{Morfill:2009}. In experiments with complex plasmas the
microparticle component, acquiring large negative charges in the plasma,
is strongly non-ideal. It is represented typically by particles of a
$\mu$m to some tens of $\mu$m in size. The internal structure of
complex plasmas is mostly determined by the mutual particle-particle
interactions consisting of isotropic (repulsive) and anisotropic
(wake-mediated) forces, as recent detailed studies demonstrated
\cite{Lampe:2000,Nefedov:2000,kroll:2010,Du:2012,Fink:2012}.
The global configuration on a large scale is determined by the
external confining forces, which are also an important structuring
element of the complex plasma dynamics
\cite{Thomas:2011,Heidemann:2011,Zhdanov:2011}. The interplay of isotropic, anisotropic and external
forces gives rise to different competing symmetries
\cite{Ludwig:2012,Woerner:2012,Zhdanov:2012}, phase
transitions
\cite{Teng:2003,Arp:2004,Melzer:2001,Wysocki:2010,Melzer:2010}, wave
patterns
\cite{Molotkov:1999,Thoma:2006,Piel:2006,Schwabe:2007,Piel:2008,Schwabe:2008,
Nosenko:2009,Menzel:2010,ThomasE:2010}, and streaming instabilities
\cite{Heidemann:2011a,Pacha:2012}, successfully observed and
explored at the kinetic level. Interdisciplinary aspects of complex plasma studies are discussed in Ref.~\cite{Ivlev:2012}.

Usually, explaining the complex plasma dynamics, it is safe to
assume that in complex plasmas the main structuring interplay is
based on the charged particles-plasma electric forces feedback
mechanism. The neutral gas in such dynamical scenarios is considered
as a passive factor, resulting mainly in the frictional reduction of
the kinetic energy of the particles. Many interesting results are known to
be explained successfully based on such simplified models, in
particular the dynamics of 2d complex plasmas (see, \emph{e.g.},
\cite{Lenaic:2012} and the references therein). Many phenomena are
also known in which all three components -- gas, dust and plasma, --
can play an equally important dynamical role as e.g. in a swirling mass
of dust forming a funnel in tornados, or gustnados
\cite{Bluestein:1999}, or in swirling dust devils which are common
in dry deserts \cite{Freier:1960}. The dust devils on Earth,
even small ones, can produce radio noise and electrical fields of
the order of 10-100 kV/m, enough to produce a breakdown in dry air
\cite{Kok:2006}.

In laboratory experiments with complex plasmas, the gas
component can be successfully used to control and manipulate the
particles enabling e.g. the rotation of the cloud as a whole under
gas drag \cite{Carstensen:2009,kroll:2010}, allowing to mimic
the magnetization of the heavy dust by making use of the complex
plasma--neutral gas frictional coupling \cite{Kaehlert:2012}, or to
study the particle cloud convection in response to gas creep
\cite{Mitic:2008,Schwabe:2011}. Using small particle clusters as a
probe of the axial discharge field distribution is also another
favorable opportunity \cite{Fink:2012}.

In this letter we report on the effects of levitating particles with
a gas flow counteracting gravity. In some sense the situation in our
plasma experiments resembles 'forced convection', in which transport of particles is induced by a gas stream
in a
number of well-known examples, \emph{e.g.}, fluidized beds
\cite{Guazzelli:2004}, snow machines, or volcano clouds. Unlike
conventional forced convection, in our experiments the
microparticles cannot be stably levitated by the gas flow alone.
The plasma influences significantly the particle dynamics. In particular,
the particle cluster stably confined inside the de Laval nozzle behaves
as a single convection 'supercell' (to some extent resembling a
microparticle blob observed in \cite{Schwabe:2009}) with the plasma
ions acting as a trigger of the marginal instability resulting in
the observed auto-wave pattern.\footnote{According to M. Faraday's publication\cite{Faraday:1831}, the term auto-wave applies to the topology of the wave structure, its stability, as well as the nature of the wave patterns.}


\section{Experimental conditions}

Experiments were conducted using a laboratory version of the
Plasmakristall-4 experimental device  designed to produce complex
plasmas inside the positive column of a high-voltage dc current-regulated discharge.
During the experiment campaign the 35-cm long and 3~cm diameter
glass discharge tube with an incorporated transparent glass de Laval
nozzle was mounted vertically. The gas inlet system and the vacuum
control system were resided correspondingly in the bottom and in the
top parts of the tube. In the experiment the input voltage was set
at $U_{dc}=1500~\text{V}$. Neon plasma at a pressure of
$100~\text{Pa}$ was sustained using a dc discharge at a
current $I_{dc} = 1.1~\text{mA}$. The electron density, electron
temperature and the electric field strength of the discharge far
away from the nozzle were estimated to be: $T_e=6~\text{eV}$,
$n_e=4\cdot10^{8}~\text{cm}^{-3}$, $E =2.1~\text{V/cm}$ (cf.
\cite{Usachev:2004,Fortov:2005, Fink:2012}).

Using the microparticle supply system located in the vicinity of the
cathode atop of the discharge tube and above the nozzle,
monodisperse melamineformaldehyde particles with a diameter of
$3.43~\mu\text{m}$ ($\pm2\%$) and a mass of $m = 3.2 \cdot
10^{-11}~\text{g}$ were injected into the dc discharge plasma. Under
those discharge conditions, as the charge estimates based on the modified Orbital Motion Limited (m-OML) model
\cite{Khrapak:2005} and the Drift Motion Limited (DML) model \cite{Morfill:2006} show, the expected microparticle
charge might be of the order of $Ze=-4000|e|$ (here $e$ is the
elementary charge).
Using the adjustable gas flow controller, a series of experiments
was performed at a fixed controllable value of the gas flow rate
($0.1-0.4~\text{sccm}$)(see footnote\footnote{Without any external modulation.
Still, it is worth noting that adjusting the flow rate excites
low-frequency oscillations along the axis of symmetry of the nozzle,
see e.g. Ref. \cite{Becker:2009}. The leading frequency $\simeq
2.5$~Hz of the flux oscillations \cite{Fink:2012} happened to be
much lower than the observed wave frequencies $\simeq 30$~Hz, see
below. The absolute error of the flow controller was estimated to be
0.006~sccm.}), powerful enough to suspend
microparticle clusters with a number density up to ~$\sim
10^6~\text{cm}^{-3}$.
The clusters formed were illuminated from the bottom using a $100~
\mu\text{m}$ wide laser sheet. The particle dynamics was recorded by a
CCD camera. The recording rate was set to 120 frames per second.

More technical details on the setup, the gas supply system, the
laser illumination system and the image recording system can be
found elsewhere \cite{Fink:2012}.

\begin{figure}
\centering\includegraphics[width=0.3\textwidth]{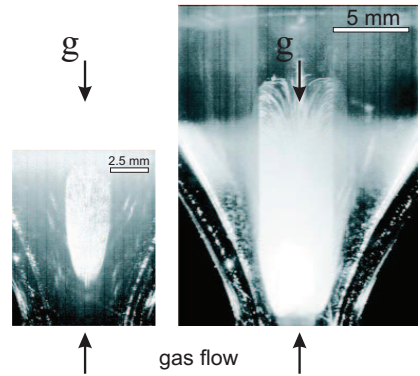}
\caption{A pellet-shaped circulation cell formed by the
microparticles deep inside the nozzle. Fifty images were superimposed
and equalized to emphasize
the shape of the nozzle walls and the
shape of the cloud. Circulation in the rarefied part of the tail of the
cloud is seen as a 'fountain stream' of separated particle
tracks. The incoming gas flow, counteracting gravity ($\mathbf{g}$),
is coming from the bottom: at 0.16~sccm, smaller cloud (left panel),
and at 0.2~sccm, bigger cloud (right panel). The scales and the
vertical positions were properly adjusted. \label{fig:1}}
\end{figure}

\section{A global design of the particle cluster}

 \begin{figure}[htbp]
\centering\includegraphics[width=0.4\textwidth]{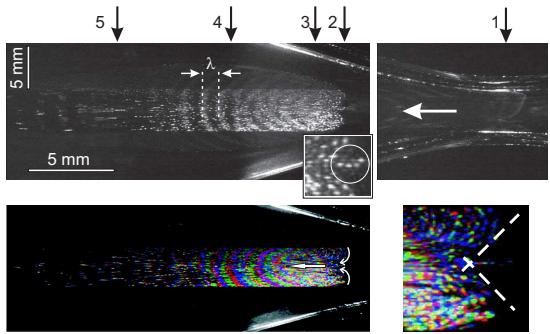}
\caption{ (Color online) Top panels: the topology of the particle cluster and the
(reconstructed) geometry of the nozzle. The main elements are: 1.
the nozzle throat; 2. the little stable satellite cloud
(highlighted by the circle in the inset); 3. the dense compact
nucleus of the main cluster; 4. the wave pattern (spread apart along
the bulk of the particle cluster); 5. the rarefied tail part (some
particle strings are well visible). The space-scale is indicated by
the white bar shown to the left. The gas flow through the nozzle
(0.4 sccm for the given experiment), counteracting the force of
gravity, is directed as a white arrow, drawn in the right image,
indicates. The acceleration of gravity in the picture is pointing to
the right, opposite the gas flow. The mean distance between the wave
ridges was measured to be $<\lambda>=(1.1\pm 0.3)~\text{mm}$.
Bottom, left: three consecutive RGB color-coded
images are superimposed to emphasize the convective particle drift
and the wave ridges propagating through.The white arrows indicate
the dominant particle convection direction. Bottom, right: the
enlarged image of the cluster nucleus with an emphasized cone-shaped
indent (the dashed lines). The cone opening angle is $2\theta
\approx 130^\circ$.
\label{fig:2}}
\end{figure}

In the course of the experiments the injected particles formed a
long-living pellet-shaped cluster, see fig.~\ref{fig:1} \footnote{Equalization is a standard function in Mathcad 14.0. This function performs a histogram equalization and is used to enhance features of a picture by spreading the pixel intensities over a larger portion of the available spectrum.}, with a
major length of $5-15~\text{mm}$ depending on their total amount of particles.
The nozzle walls, being overall transparent, are seen in this figure
as a converging funnel-shaped  set of bright laser light
reflections. The illuminating laser light comes from below, partly
scattered by entering the interior of the glass nozzle or
reflected by the nozzle walls. A small cluster (as, e.g., shown
to the left in fig.~\ref{fig:1}) with a transverse size less than
the nozzle aperture is seen completely in transmitted light; the
periphery parts of a bigger cluster are somewhat obscured (right
panel in fig.~\ref{fig:1}). Still, some particle tracks on the
cluster periphery are accessible to 
study\footnote{\url{http://iopscience.iop.org/0295-5075/102/4/45001/media} contains a movie with a spatial resolution of $33~\mu \text{m/px}$ (vertical) by $66~\mu \text{m/px} $ (horizontal). The real size of the field of view is $21.1\times 15.8~\text{mm}^2$
($\text{vertical}\times \text{horizontal}$). The $240$ frames recorded at $120~\text{fps}$ are converted into a MPEG-4 movie at a frame rate of $15~\text{fps}$.}.

The microparticle cluster, being globally of a quite simple shape,
has a complex internal structure, having quite the
appearance of a comet. The latter is known to
consist of a nucleus, coma and tail, and our cloud
(see fig.~\ref{fig:2}) also consists of a dense nucleus (unlike the
comet, it is not solid, though), a bulk with a
dancing dust making a rhythmic pattern, and a straight extended
tail. All these structural elements are dynamically very delicately
adjusted. Some of them, for instance, a cloud deformation caused by
the gas head, a little satellite cloud, and the
wave pattern, have a relatively simple explanation. The clusters were in an active,
marginally unstable state.\footnote{Such a state can lead to self-excitations in which natural modes can neither grow without bound nor decay to zero, see, \emph{e.g.},\cite{Gutowitz:2005}. } It was also
noticeable that this complicated internal design took the same
pattern for any of the explored gas flow rates.

\section{An indentation and a satellite cloud}

In the apex of the cluster a crater was dug out
by the gas head. It contains a little "satellite"
cloud, i.e. a spatially separated compact particle cloud in close proximity to the main cluster. This cloud was always inside of the cone for the whole considered parameter range. The indentation cone and the satellite cloud are
stable, at least structurally\footnote{That is
permanently restoring themselves approximately in the same shape.}.

\begin{figure}[t]
\centering\includegraphics[width=0.35\textwidth]{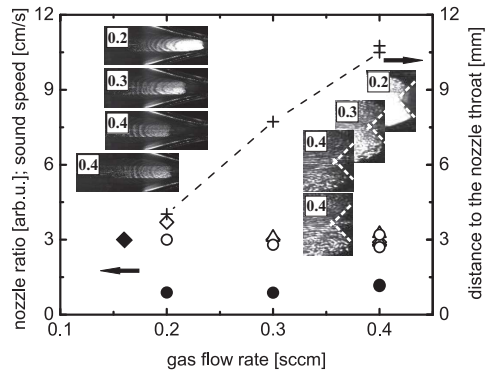}
\caption{ The 'nozzle ratio' (solid dots) and the dust sound speed
obtained with the same particle cluster for different gas flow
rates. The speed of dust sound was measured by using the density
fluctuations (open dots), the correlation methods (open triangles),
and the traced particle tracks (open diamonds). The speed of sound
obtained with a smaller cluster (fig.~\ref{fig:1}, left panel) is
shown for comparison (solid diamond). The stacked snapshots of the
cluster (FoV$\cong 1.6\times 2.1$~cm$^2$) are shown as insets to the
left. To the right are shown the enlarged images of the cluster nuclei ordered in the same way. The imposed white dashed lines indicate the same cone opening angle as that in fig.~\ref{fig:2}. The numbers give the gas flow rates measured in standard cubic centimeter per minute.
 The (nearly) constant nozzle ratio
$\eta=1.0\pm 0.3$ indicates that the stronger gas
flow pushes the entire cluster upwards away from the nozzle throat
(the crosses and the dashed line).
\label{fig:3}}
\end{figure}

The indentation cone is clearly seen in figs.~\ref{fig:2},
\ref{fig:3}. The cone size is small ($\text{cone height}/\text{cone
radius at the base} \approx \frac12$, the cone height is $\simeq
0.5$~mm). The satellite cloud is even smaller, and hence can serve
as a convenient benchmark to study the cluster dynamics. The
position $z_s$ of the center of mass of the satellite cloud can be
obtained directly using the recorded cluster images. Then,
suggesting that the axial forces are balanced, and making use of eq.~(3) in Ref.~\cite{Fink:2012}, the 'effective' gravity was calculated by the
force balance:
\begin{equation}
mg^{z}_{eff}(z_s)\equiv
mg+F^{z}_E(z_s)=m\gamma_{Eps}V^{z}_{gas}(z_s). \label{Satellite}
\end{equation}
Here $m$ is the particle mass, $g$ is the free-fall acceleration on
earth, $V^{z}_{gas}(z_s)$ is the axial gas flow velocity,
$F^{z}_E(z_s)$ is the axial projection of the resultant electric
force (effectively including e.g. the ion drag force) at the height
$z=z_s$ above the nozzle throat, and $\gamma_{Eps}$ is the Epstein
drag force coefficient \cite{Epstein:1924}, in our conditions
$\gamma_{Eps}=227$s$^{-1}$.

The result of the calculations is shown in fig.~\ref{fig:3} as a
'nozzle ratio' introduced by the relationship
\begin{equation}
\eta=\frac{g_{eff}}{g}\frac{R^2}{j}, \label{NozzleRatio}
\end{equation}
where $R[cm]$ is the nozzle transverse radius taken on at the
satellite position and $j[sccm]$ is the gas flow rate.

\begin{figure}[t]
\centering\includegraphics[width=0.35\textwidth]{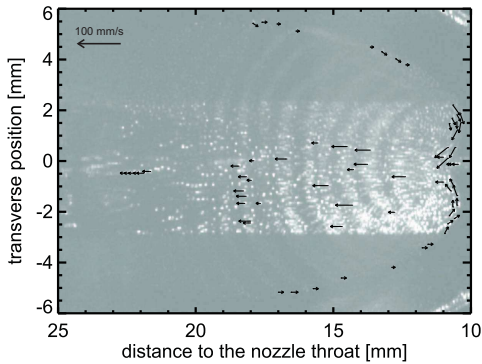}
\caption{ A single-frame velocity vector field plot of the cycling
particles at 0.4~sccm.  A few selected particle positions are shown
to emphasize the main drift directions: a relatively weak rightward
periphery drift is replaced with a rather fast particle rotation
located inside of the cluster head, which, in turn, is replaced by
the central-part leftward drift, completing the circulation loop.
The image of the particle cluster is shown in the background.
\label{fig:4}}
\end{figure}

The indentation surface is limited by the tracks of the particles
gliding along it. Since the cluster above the surface is dense, the
internal forces of the mutual interparticle interactions cannot be
neglected. Let us consider a simple hydrodynamic model of a 'liquid
with pressure', introducing a pressure force (per particle) by
$\mathbf{F}_p=-n^{-1}\nabla p$, where $n$ is the cluster density,
and $p$ is the pressure. At the surface the longitudinal component of the pressure
force has to be completely balanced but the longitudinal velocity
component is not zero, and, vise versa, transversely to the surface
the pressure force is not zero but the transverse velocity has to
vanish by definition.
Giving the small size of the indentation, it is reasonable to assume
that eq.~(\ref{Satellite}) is valid also along the surface at the position of the
satellite. Then it follows approximately that\footnote{The
transverse ($\bot$-labeled) and the longitudinal ($\|$-labeled)
components of any vector-field $\mathbf{A}$ are introduced through
its radial $r$ and axial $z$ projections by the relationships:
$A^{\bot}=A^z\sin\theta+A^r\cos\theta$,
$A^{\|}=A^z\cos\theta-A^r\sin\theta$.}
\begin{equation}
F^{\bot}_p\simeq m\gamma_{Eps}V^{\|}_d \cot\theta, \label{pressure}
\end{equation}
where $V^{\|}_d $ is the particle drift velocity and $\theta$ is the
indentation cone angle. Relationship~(\ref{pressure}) has a rather
simple physical meaning: In the framework of our model loss of
mechanical energy due to friction is to be restored by work of
pressure. Note also that in our conditions when $V^{\|}_d =
2-3~\text{cm/s}$ (see fig.~\ref{fig:4}) the pressure forces are not
small, about $F^{\bot}_p/mg=(20-30)\%$. The forces necessary to accelerate the particles in the waves are basically of the same order.

\section{A wave pattern and an instability driving mechanism}

The structure, accelerating the particles, is located in the
cluster nucleus only $(0.5-1)~\text{mm}$ apart from the center of
the satellite cloud where the particles are nearly immobile. It
accelerates the microparticles up to the velocity
$(2-3)~\text{cm/s}$ practically from rest. Note that in our
conditions the electron screening length $\lambda_{De}\simeq
1~\text{mm}$ is of the order of the electron mean free path
$\lambda_e\simeq 1~\text{mm}$ (Ne at $100~\text{Pa}$), that is of the size of
the accelerating structure. Despite the expected increase of the
plasma density deep inside the nozzle, a significant change in the screening length is not expected  because the mean kinetic
temperature of the electrons increases in the discharge constriction
as well \cite{Godyak:1988,Sirghi:1997}. The electrons are not in
equilibrium with the electric field \cite{Godyak:1988}.

The stable entrainment of particles in the upward gas flow is replaced
then by a strong oscillatory motion. Qualitatively, the main stages,
repeating in cycle, are as follows: a local sharp pulsed particle
acceleration, then a drift directed upwards, momentary stop,
short-time backward motion, etc. followed by the formation and
downstream (to gas flow) propagation of the wave ridges well
recognizable even in the raw cluster images (see fig.~\ref{fig:2}).
To help recognizing the wave dynamics further, the cluster images
were color-coded in the following manner. Three consecutive images
were taken from the sequence, and then to every of them a RGB scale
was attributed.\footnote{All pixel intensities of the first image
were treated as a scaled red color, of the second one -- as green,
and, finally, of the third -- as blue.} Afterwards they were combined in
one RGB-image shown in the bottom panel of fig.~\ref{fig:2}. Note
that the white color emphasizes the equally bright 'immobile'
pixels, e.g. those at the image edges where the 'immobile' light
reflections are located. Whereas the 'mobile' pixels have taken the
color red, green, or blue, performing a colored map.


The propagation of the strong regular wave ridges is accompanied by
the excitation of noise easily identified as cluster density
fluctuations. The results of the different noise spectra
measurements (e.g. the analysis of the density fluctuation
\cite{Menzel:2010,Heidemann:2011} or the direct particle tracking
methods \cite{Zhdanov:2010}) agree reasonably well with each other,
and with propagation analysis of the wave ridges \cite{Schwabe:2008}
(see fig.~\ref{fig:3}). The accuracy of these methods is low in our
conditions, about $30\%$. On average the sound speed was found to be $<C>=3.0\pm 1.0$~cm/s. The dust
sound speed $C$ of a dense complex plasma can be estimated by
using the relationship
\begin{equation}
C=\sqrt{\frac{ZT_i}{m}},
\label{SoundSpeed}
\end{equation}
where $T_i$ is the ion temperature, $m$ is the microparticle mass,
and $Z$ is the microparticle charge number. For our set of
parameters listed above $Z=4000$, $m = 3.2 \cdot 10^{-11}~\text{g}$,
assuming the standard values $T_i=T_{gas}=T_{room}\simeq 0.03~\text{eV}$,
it yields $C\approx 2.4~\text{cm/s}$ in a fairly good agreement with
our observations.

As in rf complex plasmas\cite{Schwabe:2007} the instability is most probably triggered by the plasma ions \cite{Shukla:1992,Rosenberg:1993,Molotkov:1999,Khrapak:2005}. It is worth noting that, albeit the pressure of 100 Pa seems
very high for waves to occur, the waves observed in our experiments
are good 'in-line' with those excited without gas flow
\cite{Khrapak:2005}, hence indicating a similarity of the excitation
mechanism.\footnote{ The instability is often originated as either the ion-dust streaming instability\cite{Shukla:1992,Khrapak:2005} or the dust-acoustic instability \cite{Rosenberg:1993,Molotkov:1999}.}
It is straightforward to find that for our set of parameters the theory predicts the following conditions for the existence of this instability:
\begin{equation}
\lambda f\simeq C, ~~~E\gtrsim E_{cr}=\frac{\gamma_{Eps}}{C}\frac{T_i}{|e|}.
\label{threshold}
\end{equation}
Both relationships are of a rather simple physical meaning. The first one merely means that the excitations are of an acoustic type, that agrees fairly well with our findings (fig.~\ref{fig:3}). The measured frequency of the wave-ridges  $<f>=29\pm 2~\text{Hz}$ as well as the wavelength $<\lambda>=1.1\pm 0.3~\text{mm}$ also correspond well to those expected. The second equation introduces an instability threshold $E_{cr}$ which is proportional to the dust damping rate, i.e. it depends on the gas pressure. It is also as intuitively expected. Giving our set of the measured parameters, we obtained $E_{cr}\approx 2.3~\text{V/cm}$, that is easy reachable inside the nozzle, also in a fairly good agreement with our observations.

\section{Convection cell}

No wave activity, at least enough intense to be traced, was recorded
inside the tail part of the cluster (fig.~\ref{fig:2}). This most
rarefied part of the cloud is dominated by the weak particle drift
directed upwards along the gas flow lines close to the nozzle axis
\footnote{The particle tracks are a bit perturbed by the wave
activity located upstream; on average, $V_{drift}=(4\pm
1)~\text{mm/s}$.} and downwards at the periphery of the cluster (see
fig.~\ref{fig:4} demonstrating the momentary velocity vector field
of the drifting particles). In fig.~\ref{fig:4} it is clear that the
particles are forming vortices at the edges of the cluster,
developing a single complete circulation cell. By analogy with
\cite{Mitic:2008,Schwabe:2009} convection, at least partly, might be
caused by a creeping gas flow. The circulation cell size somewhat
varies with the gas flow rate. There also might be some variation
with height. However, it cannot to be studied in detail in this
experiment setup because the parts in the very periphery are
poorly accessible via the applied laser illumination. A few traced
particles moving back down along the outside of the cloud are shown
in fig.~\ref{fig:4}.

Convection of the particles, a clearly
recognizable feature of their motion close to the nozzle axis, can affect the low
frequency oscillations but cannot be the reason for the
strong wave excitations, merely because its maximum frequency is
much lower than that of the wave-ridges. The maximum angular velocity of conveying particles in the head
of the cluster is $\omega_c\approx 20-30~\text{s}^{-1}$, that is 6-7 times lower than that measured for the waves.

\section{Discussion}

In the present work, using a nozzle-shaped glass insert in the glass tube plasma chamber of the PK-4 experiment and a gas flow for suspending the particles against gravity, a complicated structure within the microparticle cloud located above the nozzle has been observed and analyzed. The main results of these investigations are: (i) There is an extended stable particle cloud confined above the nozzle showing convection as well as an autowave pattern; (ii) This complicated and stable structure is the result of the various internal and external forces acting on the microparticles, i.e. gravity, longitudinal and radial electric fields in the presence of the nozzle, an applied gas flow from below, the Coulomb interaction between the microparticles, and the interaction of the microparticles with the ions; (iii) The global structure of the main cloud was explained quantitatively by employing the balance of the various forces acting on the particles; (iv) A possible explanation for the convection observed within the main cloud could be creeping gas flow along the glass walls of the chamber and nozzle as also seen in previous investigations with PK-4; (v) The autowaves or self-excited waves observed in the main cloud result from the well-known ion-dust streaming instability observed in complex plasmas under various conditions.

\section{Conclusion}

To summarize, we have reported a direct observation of a wave pattern
disclosed by dense microparticle clusters stably confined in a dc
discharge above a de Laval nozzle.
The clusters were in an active,
marginally unstable state, accompanied by wave excitations. The wave
pattern consisted of steep regular wave ridges, fast propagating
through the cloud, as well as of low amplitude density fluctuations,
originated as dust sound. It is noticeable that varying the
cluster position by adjusting the gas flow had no or only little
effect on the wave propagation speed. We have argued that the
instability is triggered by the plasma ions by means of a well-known
ion-dust streaming instability mechanism. The measured space and times
scales of the waves are in fairly good agreement with this theory,
which supports this claim.

\section{Acknowledgements}

The research leading to these results has received funding from the
European Research Council under the European Union's Seventh
Framework Programme (FP7/2007- 2013)/ERC Grant Agreement No. 267499,
as well as funding by a Marie Curie International Outgoing
Fellowship within the 7th European Community Framework Programme,
and by DLR (BMBF) under Grants No. 50WP0204, No. 50WM0504, No.
50WM0804, and No. 50WM1150. The authors would like to thank Ch. Rau
for his help.

\end{document}